\documentclass[10 pt]{article}
\begin{document}

\baselineskip 18 pt

\author{A.Y. Shiekh\footnote{\rm shiekh@dinecollege.edu} \\
{\it Din\'{e} College, Tsaile, Arizona, U.S.A.}}
\title{{\bf The role of Quantum Interference in Quantum Computing}}
\date{August 2005}
\maketitle

\begin{abstract}
Quantum interference is proposed as a tool to augment Quantum Computation.
\end{abstract}

\section{Can quantum interference augment Quantum Computation?}

\subsection{The dilemma}

The power of quantum computing lies in the quantum ability to have a linear superposition
of all possible states (Hadamard spread), so that one could produce the following, non-entangled, 3
qubit state (we ignore the normalizations throughout for simplicity).
\begin{eqnarray*}
\left| \psi  \right> &=&
(\left| 0 \right> + \left| 1 \right>)
(\left| 0 \right> + \left| 1 \right>)
(\left| 0 \right> + \left| 1 \right>) \\ &=&
\left| 0 0 0 \right> + 
\left| 0 0 1 \right> +
\left| 0 1 0 \right> +
\left| 0 1 1 \right> + \ldots +
\left| 1 1 1 \right>
\end{eqnarray*}
A function $f$ applied to this one state results in the evaluation of this function for each
component, and herein lies the exponential parallelism of quantum computing.

This information, however, is not directly accessible, as the act of measurement picks out only one
of the components, and therein lies the dilemma of quantum computing.

\subsection{The response}

Let us start with the same state as above, and like Grover's algorithm, apply the decision
function to mark the invalid solutions by changing their sign. For the sake of argument let us
suppose that solutions 001 and 010 satisfy the function, which yields the state:
$$
   - \left| 0 0 0 \right> + 
{\bf \left| 0 0 1 \right>}
   - \left| 0 1 0 \right> +
{\bf \left| 0 1 1 \right>} + \ldots
   - \left| 1 1 1 \right>
$$
which has got us nowhere at all, {\it unless} one were to bring in the mechanism of Young's double
slit or the beam splitter interferometer, with the marking function being applied to one of the two
arms alone. Then interference would yield:
\begin{eqnarray*}
 &-& \left| 0 0 0 \right> + 
{\bf \left| 0 0 1 \right>}
   - \left| 0 1 0 \right> +
{\bf \left| 0 1 1 \right>} - \dots
   - \left| 1 1 1 \right>
\\
 &+& \left| 0 0 0 \right> + 
{\bf \left| 0 0 1 \right>} +
     \left| 0 1 0 \right> +
{\bf \left| 0 1 1 \right>} + \ldots +
     \left| 1 1 1 \right>
\end{eqnarray*}
to expose the desired solutions
$$
{\bf \left| 0 0 1 \right>} +
{\bf \left| 0 1 1 \right>}
$$
one of which will be seen upon measurement, and can be confirmed on a classical computer, if so
desired. The two arms are brought into overlap and not sent through a final beam splitter as is
typical of an interferometer.

To locate the remaining solution, one can start over, and exclude the known solution by also
flipping its sign in one of the two interference arms. Eventually all solutions will be located and
removed, so the final run will expose either a non-valid solution or a previously found solution
from the remnants of the wave-function.

Concerns over lost unitarity can be allayed by noting that a quantum computer typically starts by
transforming a sharp (ground) state into a superposition, and that this is a unitary change. All
that is happening here is the inverse, and so the process is also unitary.

In practice, due to imperfect cancellation, this process may need to be repeated a few times.

\section{Implications for error correction}

The amplitudes are analogue and the states digital in nature; so errors in the amplitude, no matter
how small, might be expected to affect all qubits in an error correction scheme, which so could not
be completely removed. In contrast, small errors in the state would only flip one qubit of a
state, which is amenable to correction.

Since the approach to quantum computing being proposed here is insensitive to small changes in
amplitude, this technique may benefit from digital robustness.

\section{Conclusion}

By explicitly introducing an interferometer, a generic exponential speedup seems possible, and the
system may be more robust to error correction.

After completion of this work it was discovered that Finkelstein and Castagnoli had also been
working on using quantum interference to achieve a generic speed up of a quantum computer, under
the name quantum interferometric computation (QUIC)\footnote{Private Communication Dec 2nd 2005:
Professor David Finkelstein}.

\section{References}

M. Nielsen, I. Chuang, {\it Quantum Computation and Quantum Information}, Cambridge University
Press, (2000).

\end{document}